\documentclass[aps,twocolumn,groupedaddress,showpacs,nofootinbib]{revtex4}

\usepackage{graphicx}
\usepackage{hyperref}
\usepackage{dcolumn}
\usepackage{bm}
\usepackage{epsfig,amsbsy,bm,amssymb,amsmath}
\usepackage{xcolor}

\newcommand{\hs}{\hspace*}
\newcommand{\vs}{\vspace*}
\newcommand{\np}{\newpage}
\newcommand{\w}{\omega}

\newcommand{\eref}[1] {(\ref{#1})}
\newcommand{\Eref}[1] {Eq.~(\ref{#1})}
\newcommand{\Fref}[1] {Fig. \ref{#1}}

\newcommand{\nn}{\nonumber}
\newcommand{\be}{\begin{equation}}
\newcommand{\ee}{\end{equation}}
\newcommand{\br}{\begin{eqnarray*}}
\newcommand{\er}{\end{eqnarray*}}
\newcommand{\ba}{\begin{eqnarray}}
\newcommand{\ea}{\end{eqnarray}}
\newcommand{\bp}{\begin{minipage}}
\newcommand{\ep}{\end{minipage}}
\newcommand{\bt}{\begin{tabular}}
\newcommand{\et}{\end{tabular}}

  \newcommand{\e}{{\bm e}}
  
 \newcommand{\ig}[1]{\includegraphics[width={#1}]}
\newcommand{\Li}{Li$^+$~}

\renewcommand{\t}{\tau}

\renewcommand{\e}{\epsilon}
  \newcommand{\ve}{\varepsilon}

\begin{document}
\bibliographystyle{apsrev}
%

\title{Fano line shape metamorphosis in resonant two-photon ionization}

\author{Vladislav ~V. Serov$^{1}$}
\author{Anatoli~S. Kheifets$^{2}$}

\affiliation{$^{1}$General, Theoretical and Computer
  Physics, Saratov State University, Saratov
  410012, Russia}

\affiliation{$^{2}$Research School of Physics, The Australian National
  University, Canberra ACT 2601, Australia}
\email{A.Kheifets@anu.edu.au}

 \date{\today}


\begin{abstract}

Two-photon atomic ionization driven by time-locked XUV and IR pulses
allows to study dynamics of Fano resonances in time and energy
domains.  Different time evolution of the two interfering pathways
leading to a Fano resonance can be exploited to turn the Fano profile
of the two-photon XUV/IR ionization into a symmetric Gaussian once the
directly ejected photoelectron leaves the parent ion and cannot any
longer absorb an IR photon.  This line shape transformation allows for
the direct determination of the resonant lifetime from the
spectroscopic measurements without need for an extremely fine energy
resolution. Ubiquitous nature of Fano resonances makes this
determination a universal tool in diverse quantum systems ranging from
nuclei to nano-fabricated solids.

\end{abstract}

\maketitle

Ugo Fano's most enduring legacy in atomic physics is his theory of
Bound states embedded Into Continuum (BIC)
\cite{Fano1935,PhysRev.124.1866,PhysRev.137.A1364}. Fano's theory
explained an earlier \cite{Silverman1964} and stimulated subsequent
\cite{Madden1963} measurements which detected distinctive asymmetric
line shapes in atomic ionization cross-sections.
The cross-section near the Fano resonance  takes the form
\be
\sigma(\e) = 
|D(\e)|^2
\propto
{(\e+q)^2\over \e^2+1}
\ \ , \ \ 
\e= {E-E_0\over \Gamma/2}
\ .
\label{Fano}
\ee
%
Here $\e$ is a detuning from the resonance center $E_0$ measured in
units of the resonance half width and $q$ is the Fano shape
index. The ionization amplitude in \Eref{Fano} can be 
expressed via the phases of the resonant and non-resonant
(background) scattering \cite{Connerade1988}:
\be
D(\e) \propto [e^{2i(\delta+\phi)}-1]/2
\ \ , \ \ 
\cot\delta=\e
\ \ , \ \ 
\cot\phi=q
\ee
In the absence of the background scattering when $\phi=0$ the Fano
profile turns into a Lorentzian which is characteristic for an
exponential decay of a discrete excited state with a finite lifetime
$\tau = 1/\Gamma$ \cite{Lorentz1916}.

The BIC phenomenon is ubiquitous in nature.  In recent decades, Fano
resonances have been observed in very diverse systems such as
M\"ossbauer nuclei \cite{Li2023}, quantum dots \cite{Kroner2008},
plasmonic nanostructures \cite{Fan2010, Luk'yanchuk2010,Rahmani2012},
2D photonic crystals \cite{Zhou2014}, metasurfaces \cite{Yang2015} and
exploited widely in nanotechnology and nanophotonics
\cite{Miroshnichenko2010,Hsu2016,Limonov2017,Koshelev2023}.

In atomic physics, a renewed interest to Fano resonances has been
stimulated by availability of ultra-short laser pulses.  These novel
laser sources have allowed to study ultrafast dynamics of Fano
resonances both in the time and energy domains
\cite{Ott2013,Gruson734,Kotur2016,Kaldun2016,Cirelli2018,Busto2018,Barreau2019,Turconi2020,Neoricic2022}. Not
only can the Fano resonances be probed with lasers. Intense laser
pulses have the capacity to induce a coupling of a resonant state to
the continuum which otherwise does not naturally occur
\cite{Litvinenko2021}.
Lasers can also blur the difference between the Fano and Lorentz line
shapes \cite{Ott2013}. The photoelectron continuum phase $\phi$ can be
offset by a ponderomotive energy kick from a strong IR pulse.  Such a
compensation turns a Fano photoabsorption line into
a Lorentzian.

Even though autoionizing two-electron excitations in He were
discovered nearly a century ago \cite{Compton1928}, their studies
still remain a very  active research area, i.e.
\cite{Lin1986,Rost1997,Argenti2008}.
Time resolution of the Fano resonances  sheds new light on 
this phenomenon by providing a universal phase control
\cite{Ott2013,Kotur2016} and monitoring the birth of a photoelectron
\cite{Gruson734,Kaldun2016}. However, there is little direct overlap
between the old and new physics of Fano resonances.
Majority of time resolved studies of Fano resonances have been
conducted using the technique of Reconstruction of Attosecond Beating
By Interference of Two-photon Transitions (RABBITT)
\cite{Gruson734,Cirelli2018,Busto2018,Barreau2019,Turconi2020,Neoricic2022}.
Even though resonant $r$RABBITT method can scan the photoelectron
spectral lines with a sufficient resolution, these lines remain too
broad for direct spectroscopic determination of the accurate resonance
position and width. This become particularly problematic for highly
excited and narrow resonant states. Neither can the Fano shape index
be measured directly from photoelectron line shapes. The bound state
structure of the target can be deduced from an under-threshold
$u$RABBITT process
\cite{SwobodaPRL2010,Villeneuve2017,PhysRevA.103.L011101,Kheifets2021Atoms,Neoricic2022}.
However, this determination is restricted to single-electron
excitations below the ionization threshold.

In the RABBITT technique, an atom is ionized by an XUV pump in the
form of  an attosecond pulse train (APT). The corresponding
photoelectron spectrum contains a comb of primary harmonic peaks
centering at the odd multiples of the IR probe frequency
$\Omega_{2q\pm1}=(2q\pm1)\w$. Arrival of an IR probe adds new spectral
features known as sidebands (SB's). They correspond to an even
frequency $\Omega_{2q}=2q\w$ and are  formed by the two competing
pathways $\Omega_{2q+1}-\w$ and $\Omega_{2q-1}+\w$.  These two
pathways interfere and the SB population oscillates as the XUV/IR
delay $\Delta$ varies:
\be
S_{\rm SB}(\Delta) \propto
b\cos[2\omega\Delta-\delta]
\ \ , \ \ 
\delta = 2\w\t_a
\ .
\label{RABBITT}
\ee
The magnitude $b$ and phase $\delta$ of the RABBITT oscillations can be
expressed via two-photon ionization amplitudes
\cite{Dahlstrom201353}. The RABBITT phase can then be converted to the
atomic time delay $\t_a=\delta/(2\w)$.

\begin{figure}
\ig{0.9\columnwidth}{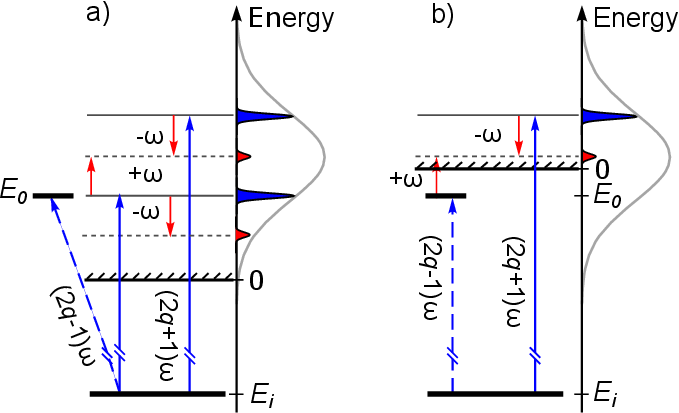}
\caption{Schematic representation of the resonant $r$RABBITT (a) and the
  under-threshold $u$RABBITT (b) processes. Direct and
  resonant XUV absorption pathways are shown with solid
  and dashed lines, respectively. 
\label{Fig1}}
\end{figure}

In the resonant $r$RABBITT process, one of the main odd harmonics is
tuned to an autoionization resonance as shown in \Fref{Fig1}a.  In
this case, the RABBITT parameters of the two neighboring SBs display a
clear resonant structure. In the underthreshold $u$RABBITT process
illustrated graphically in \Fref{Fig1}b, the discrete excited state is
located below the threshold and only the lowest SB is affected by the
resonance.
In principle, an $r$RABBITT measurement can detect the lifetime of the
resonance. However, because of the $2\w$ periodicity in
\Eref{RABBITT}, the useful range of the XUV/IR delays is restricted to
the half cycle of the IR oscillation which is 2.6~fs for the commonly
used 800~nm wavelength. This range is far too narrow for majority of atomic
and molecular autoionizing resonances.

In the present work, we remove this restriction by employing an isolated
XUV pulse instead of an APT.
%
%
We assume that the duration of the XUV pulse is much shorter than the
lifetime of the autoionizing state $\tau$. In the meantime, its
spectral width is smaller than $\w$ such that SB's are well separated
from the main photoelectron line. We consider an IR pulse of the
duration $T$ with a Gaussian envelope
\be
E(t)=f(t)\sin[\omega(t-\Delta)]
\  , \  
f(t)=\exp\left[-\frac{(t-t_0)^2}{2T^2}\right]
 .
\ee
%
%
Arrival of the IR pulse
populates the two SB's centered at $E_0 \pm \omega$. The SB amplitude
is obtained by a simple Fourier transform
\be A_{\rm SB}(\ve) \sim \int_0^\infty \!\!\!  \exp\left[i\ve t -
  {(t-t_0)^2\over 2T^2}-{t\over 2\tau}\right] dt \ ,
\label{integral}
\ee
%
%
where $\ve$ is the absolute detuning from the center of the SB. The
integration in \Eref{integral} is carried out from the moment of
arrival of the XUV pulse at $t=0$, when the autoionizing state is
instantly (compared to $T$ and $\tau$) populated. The integration
results in
\ba \nn A_{\rm SB}(\ve) &\sim& \exp \left[-
  \frac{T^2(2i\ve\tau-1)^2}{8\tau^2} - i\ve t_0 - \frac{t_0}{2\tau}\right]
\\ &\times& \left[\mathrm{erf}\left\{\frac{T(2i\ve\tau-1)}
  {\sqrt8\tau}+\frac{t_0}{\sqrt2T}\right\}+1\right] 
\ .
\label{gauss}
\ea
%
At $t_0 \gg T$  the SB acquires a  Gaussian line shape
\be S_{\rm SB}(\ve)=|A_{\rm SB}(\ve)|^2 \sim \exp\left[- T^2\ve^2 -
  \frac{t_0}{\tau}\right] 
\ee
%
%
The width of the Gaussian is determined by the length of the IR pulse
$T$ but not the width of the resonance $\Gamma$. The magnitude of the
Gaussian decreases exponentially with increasing $t_0$. The timing
constant of this exponential decay is equal to the lifetime of the
autoionizing state $\tau$.

Qualitatively, the line shape \eref{gauss} is easy to understand.  The
IR pulse ejects into the continuum a number of electrons proportional
to the population the autoionizing state at the moment of its
arrival. If $t_0=0$ and $T \gg \tau$, the distribution \eref{gauss}
becomes a Lorentzian
\be
 S_{\rm SB}(\ve)|^2 \sim (\ve^2+\tau^2/4)^{-1}
\ee 
%
%
In addition to exciting an autoionizing state (resonant excitation,
dashed line in \Fref{Fig1}a), the XUV pulse causes the direct
ionization (solid line in the same figure).  Thus the resonant SB
amplitude is the sum of the two terms
\be
A_{\rm SB}(\ve) = A_{\rm SB, R}(\ve) + A_{\rm SB, CC}(\ve) \ . 
\ee
%
%
Here $A_{\rm SB, CC}(\ve)$ is the amplitude of the continuum-continuum
transition in the IR field. At $t_0=0$, this amplitude has a
magnitude comparable to $A_{\rm SB, R}(\ve)$, so their summation
leads to an asymmetric line shape similar to the Fano profile.  The
wave packet of electrons ejected directly by the XUV pulse moves away
from the ion as $t$ grows. At sufficiently long XUV/IR delay
$t_0\to\infty$, the IR pulse will not alter the energy of the
photoelectron. Classically, a free electron cannot absorb a photon to
conserve both the momentum and energy. Hence, in the case of a large
XUV/IR delay, $A_{\rm SB, CC}(\ve,t_0\to\infty) \to 0$ so
that the SBs acquire a symmetric Gaussian line shape.

\begin{figure}
\ig{0.8\columnwidth}{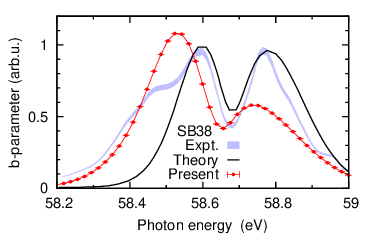}

\ig{0.8\columnwidth}{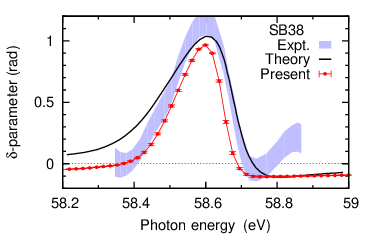}

\ig{0.8\columnwidth}{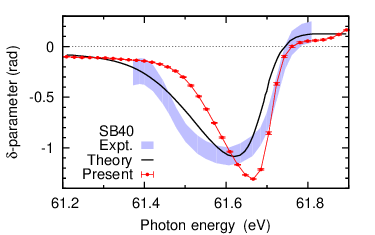}
\caption{ $r$RABBITT magnitude and phase parameters in He when H39 is
  tuned to the $sp2^+$ resonance. Top: the magnitude $b$ parameter of
  SB38.  Middle and bottom: the phase $\delta$ parameters of SB38 and
  SB40, respectively. The experimental and theoretical results of
  \cite{Busto2018} are compared with the present calculation. The
  error bars in the latter indicate the accuracy of the fit with
  \Eref{RABBITT}.
\label{Fig2}
\vs{-0.3cm}}
\end{figure}

This simple analytical model is supported by accurate numerical
simulations in two-electron targets, the He atom and the \Li ion.  In these
simulations, we solve numerically the time-dependent Schr\"odinger
equation (TDSE) on a multi-configuration basis as described
in detail in \cite{Serov2024}. 
To test and validate our numerical technique, we first conduct the
$r$RABBITT calculation on He and make a comparison with experimental
and theoretical results presented in an earlier work
\cite{Busto2018}. This comparison is shown in \Fref{Fig2}. Here we
display the RABBITT magnitude $b$ and phase $\delta$ parameters in
\Eref{RABBITT} for SB38 and SB40 when the harmonic H39 is tuned to the
$sp2^+$ resonance.  The perturbation theory calculation in
\cite{Busto2018} and the present numerical results sandwich the
experimental data sufficiently closely. From this we can conclude that
our numerical accuracy is at least not worse than that in the previous
work \cite{Busto2018}.

\begin{figure}
\ig{0.8\columnwidth}{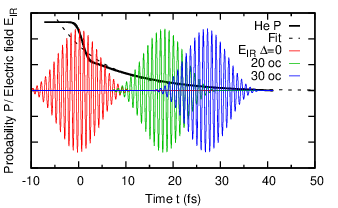}

\ig{0.8\columnwidth}{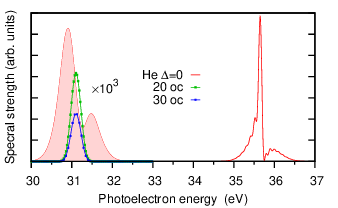}

\hs{-5mm}
\ig{0.85\columnwidth}{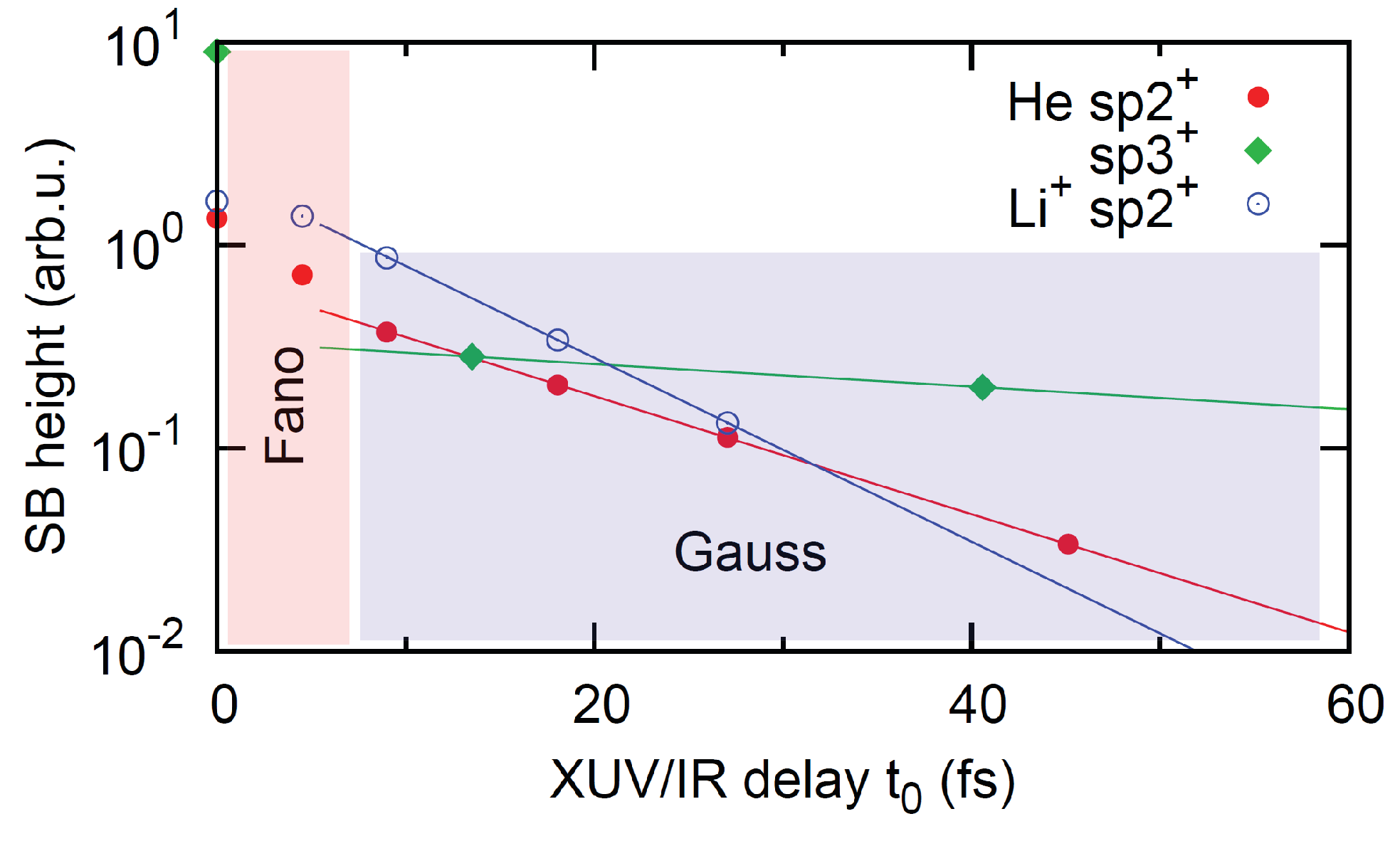}
\caption{Top: the thick solid line shows the scaled probability of
  locating the departing photoelectron within the simulation
  boundary. Differently colored IR probes arrive at various delays (in
  units of optical cycles, 1~oc=0.9~fs). Middle: the photoelectron
  spectrum corresponding different XUV/IR delays. The SBs (only one
  set is shown) are scaled by a factor $10^3$ for better
  visibility. Bottom: the SB height as a function of the XUV/IR delay
  is fitted with an exponential decay function. The shaded areas mark
  the Fano and Gauss line shape appearance. The top and middle panels
  display the He $sp2^+$ data whereas the bottom panel shows the He
  $sp2^+, sp3^+$ and Li$^+$ $sp2^+$ data.
\label{Fig3}}
\end{figure}

Once our numerical method is tested and validated, we return to the
core problem of the present investigation.  To do so, we substitute
the APT in $r$RABBITT with an isolated XUV pulse. In addition, we
shorten the duration of the IR pulse from 21~fs in RABBITT simulations
to 7~fs in the Fano line shape study. We consider the same lowest
two-electron excitation $sp2^+$ in He. Of all the helium autoionizing
states, this resonance has the shortest lifetime of 17~fs which
corresponds to $\Gamma=36$~meV \cite{Domke1996}. This resonance was
not considered in \cite{Ott2013} because the IR pulse could couple it
strongly with non-dipole $2p^2$ excitations \cite{Loh2008,Chu2011}. We
eliminate this complication by tripling the energy of the IR photon
from 1.55~eV (800~nm) as in \cite{Busto2018} to 4.6~eV (266~nm).

The graphical illustration of our numerical results is shown in
\Fref{Fig3}. The XUV probe arrives at the time zero and sets in motion
the photoelectron which leaks outside the simulation boundary. The
thick solid line in the top panel of the figure shows the probability
of finding the photoelectron inside this boundary.  Once this
probability decreases with time, the IR probe pulse arrives at
variable delay. This probe adds two SBs to the main peak in the
photoelectron spectrum. Only one set of SB's is shown in the middle
panel of the figure for better clarity. The main peak exhibits a
typical asymmetric Fano line shape irrespective of the XUV/IR
delay. When the XUV pump and IR probe overlap, the SBs display similar
asymmetric line shapes.  As the XUV/IR delay $\Delta$ grows and the
photoelectron population inside the boundary starts to decrease, this
asymmetry of the SB vanishes and the resonance decays exponentially as
prescribed by \Eref{gauss}. From this point on, by monitoring the SB
height, we can deduce the lifetime of the resonance. This procedure is
illustrated in the bottom panel of the figure where the SB height is
fitted with an exponential decay function
$\propto\exp(-\Delta/\tau)$. The lifetime of the resonance $\tau=
15$~fs found this way is rather close to the experimental value of
17~fs \cite{Domke1996}.  We repeat this determination for other
autoionizing states. For the $sp3^+$ resonance of He, we need not
increase the IR frequency as it does not mix with other autoionizing
states. Our determination returns the lifetime values 
of 81.7~fs (upper SB) and 79.5~fs (lower SB). This is to compare with
the experimental value of 82~fs \cite{Domke1996}.
The $sp2^+$ resonance in Li$^+$ ion has the line
width of 74~meV and the corresponding life time of 8.7~fs
\cite{Carroll1977}. Our determination as illustrated in the bottom
panel of \Fref{Fig3} returns $\tau=9.6$~fs. 

Incidentally, the exponential decay fit of the autoionizing state
population as shown in the top panel of \Fref{Fig3} returns identical
$\tau$ values for both targets. So the difference between the
theoretical and experimental lifetimes is due to deficiency of the
theoretical description of the resonance with a limited number of
configurations. The exponential fitting of the SB's does not introduce
any error and once applied in experiment will return the accurate
lifetime.

\np In conclusion, we study ultrafast dynamics of the autoionizing
states in two-electron atomic targets driven by a combination of an
ionizing XUV pump pulse and a delayed and weak IR probe pulse. The
two-photon ionization manifests itself by appearance of sidebands
displaced from the main photoelectron line by the energy of the IR
photon. When the XUV pump is tuned to the resonance, the main peak in
the photoelectron spectrum acquires a characteristic asymmetric Fano
line shape. In the meantime, the shape of the sidebands varies
depending on the XUV/IR delay. When this delay is sufficiently large
and the directly ionized electron leaves the parent ion, the Fano line
shape of the sidebands turns into a symmetric Gaussian. In this
regime, a free electron cannot absorb an IR photon and the direct
pathway cannot compete with its resonant counterpart.  As the XUV/IR
delay is increased further, evolution of the SB height becomes
particularly simple and can be represented by an exponential decay
function. The use of this function allows for an accurate
determination of the lifetime of the resonance. This behavior is
predicted analytically and further supported by accurate numerical
simulations. Even though the presented calculations are performed for
atomic targets, our development of the molecular TDSE code
\cite{Serov2024} allows for similar simulations in molecules such as
H$_2$. 

The proposed determination of the resonant life time does not require
an extremely fine spectral resolution. It is sufficient for this
determination that the SB height decreases with XUV/IR delay
while the resonance is not fully spectrally resolved.  This is
particularly beneficial for studying highly excited and long living
autoionizing states in atoms and molecules while the photoelectron
energy resolution is modest. This is in contrast to other
spectroscopic techniques. For instance, transient photoabsorption
spectroscopy of the He autoionizing states with a lifetime exceeding
100~fs requires a spectral resolution better than 20~meV
\cite{Ott2013}.  The more severe case is a M\"ossbauer nuclei which
exhibits resonances with lifetime exceeding 100~ns and the spectral
width of the order of neV \cite{Li2023}. For such narrow resonances
the proposed technique could be particularly beneficial since its
accuracy grows for longer lifetimes exceeding duration of the probe
pulse. Because Fano resonances are so prevalent in nature, we expect a
wide application of the proposed method. The spectral range of the
pump and probe pulses can vary from THz radiation in semiconductors
\cite{Litvinenko2021} to $\gamma$-rays in nuclei
\cite{Li2023}. Nevertheless, the basic principle of the proposed
technique remains the same. 

\paragraph*{Acknowledgment:}  
This work was supported by the Discovery Grant DP190101145 from the
Australian Research Council.


\end{document}